\documentclass[letterpaper,twocolumn,10pt]{article}
\usepackage{usenix,epsfig}
\usepackage[hyphens]{url}
\usepackage[dvipsnames]{xcolor}
\usepackage[compact]{titlesec}
\usepackage{hyperref}
\hypersetup{
    colorlinks,
    citecolor=black,
    filecolor=black,
    linkcolor=black,
    urlcolor=black
}
\usepackage[style=ieee,natbib=true,sorting=none]{biblatex} 
\usepackage{amsmath}
\usepackage{comment}
\usepackage{relsize}
\usepackage[capitalize]{cleveref}
\usepackage{listings}
\crefname{lstlisting}{listing}{listings}
\Crefname{lstlisting}{Listing}{Listings}
\crefname{section}{Sec.}{Sec.}
\Crefname{section}{Sec.}{Sec.}
\usepackage{inconsolata}
\usepackage{placeins}
\usepackage{standalone}

\usepackage{tikz}
\usetikzlibrary{arrows.meta,positioning,shapes,shadows,fit,calc,backgrounds}
\usepackage{tikz-network}

\usepackage{float}
\usepackage{subcaption}

\usepackage{enumitem} 
\usepackage{textcomp} 
\usepackage{annotate-equations}

\usepackage{authoraftertitle}

\newcommand{\beginsupplement}{%
        \setcounter{table}{0}
        \renewcommand{\thetable}{S\arabic{table}}%
        \setcounter{figure}{0}
        \renewcommand{\thefigure}{S\arabic{figure}}%
        \setcounter{section}{0}
        \renewcommand{\thesection}{S\arabic{section}}%
        \setcounter{lstlisting}{0}
        \renewcommand{\thelstlisting}{S\arabic{lstlisting}}%
        \twocolumn[
         \vbox to 1.0in{
         \vspace*{\fill}
         \vskip 2em
         \begin{center}%
          {\Large\bf \MyTitle \par}
          {\large\bf Supplementary Material \par}
          \vskip 0.375in minus 0.300in
         \end{center}%
         \par
         \vspace*{\fill}
          \vskip 1.5em
         }
        ]
}

\begin{document}

\makeatletter
\g@addto@macro\@maketitle{
\vspace{2.5cm}
\begin{figure}[H]
\setlength{\linewidth}{\textwidth}
\setlength{\hsize}{\textwidth}
\centering
    \begin{subfigure}[b]{0.3\textwidth}
        \centering
        \begin{tikzpicture}[scale=0.8]
\SetDistanceScale{1.8}
\Plane[x=1.0,y=1.0,width=3,height=3.0]
\Vertices[shape=circle,NoLabel=True]{centralization-figure/powerlaw_vertices_flat.csv}
\Edges[color=black]{centralization-figure/powerlaw_edges_flat.csv}
\draw[draw=black, fill=white] (1.95,5.98) rectangle ++(2.2,1.05);
\Vertex[label=Community,label=Community,position=right,size=0.25,color=orange,x=1.25,y=3.70] {C};
\Vertex[label=Community,label=User,position=right,size=0.2,color=cyan,x=1.25,y=3.48] {U};
\end{tikzpicture}
        \vspace*{-10mm}
        \caption{Centralized}
        \label{fig:centralization-pl}
    \end{subfigure}
    \hfill
    \begin{subfigure}[b]{0.3\textwidth}
        \centering
        \begin{tikzpicture}[scale=0.8]
\SetDistanceScale{1.8}
\Plane[x=0.7,y=0.5,width=3,height=3.0]
\Vertices[shape=circle,NoLabel=True]{centralization-figure/er_vertices_flat.csv}
\Edges[color=black]{centralization-figure/er_edges_flat.csv}
\end{tikzpicture}
        \vspace*{-10mm}
        \caption{Decentralized}
        \label{fig:centralization-er}
    \end{subfigure}
    \hfill
    \begin{subfigure}[b]{0.3\textwidth}
        \centering
        \begin{tikzpicture}[scale=0.8]
\SetDistanceScale{1.2}
\Plane[x=0.8,y=0.8,width=4.50,height=4.50]
\Vertices[shape=circle,NoLabel=True]{centralization-figure/hybrid_vertices_flat.csv}
\Edges[color=black]{centralization-figure/hybrid_edges_flat.csv}
\end{tikzpicture}
        \vspace*{-10mm}
        \caption{Ambiguous}
        \label{fig:centralization-mixed}
    \end{subfigure}
    \caption{The influence of a community is tied to both its size and topological role in a network. In the centralized network, the orange community at the center both has the largest population of blue users, and serves as a bridge between four other communities. In the decentralized example, communities are of variable size, but none have a pivotal position to influence their peers. In the ambiguous case, one community is much larger, but the remaining network matches the ``decentralized" example. Neither a distribution of community sizes nor purely structural measurements like betweenness centrality or graph conductance adequately capture this notion of community-level influence.}
\label{fig:centralization}
\end{figure}
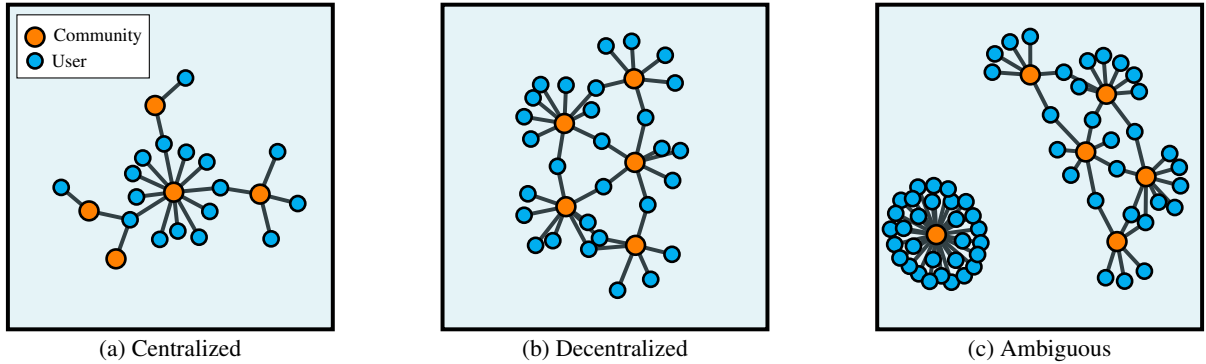
}
\makeatother

\date{}

\title{Measuring Centralization of Online Platforms Through Size and Interconnection of Communities}

\author{
{\rm Milo Z.\ Trujillo}\\
University of Vermont \\
\texttt{milo.trujillo@uvm.edu}
\and
{\rm Laurent H\'{e}bert-Dufresne}\\
University of Vermont \\
\texttt{laurent.hebert-dufresne@uvm.edu}
\and
{\rm James Bagrow}\\
University of Vermont \\
\texttt{james.bagrow@uvm.edu}
} 


\maketitle

\begin{abstract}
Decentralized architecture offers a robust and flexible structure for online platforms, since centralized moderation and computation can be easy to disrupt with targeted attacks.
However, a platform offering a decentralized architecture does not guarantee that users will use it in a decentralized way, and measuring the centralization of socio-technical networks is not an easy task.
In this paper we introduce a method of characterizing community influence in terms of how many edges 
between communities would be disrupted by a community's removal.
Our approach provides a careful definition of ``centralization" appropriate in bipartite user-community socio-technical networks, and demonstrates the inadequacy of more trivial methods for interrogating centralization such as examining the distribution of community sizes.
We use this method to compare the structure of multiple socio-technical platforms -- Mastodon, git code hosting servers, BitChute, Usenet, and Voat -- and find a range of structures, from interconnected but decentralized git servers to an effectively centralized use of Mastodon servers, as well as multiscale hybrid network structures of disconnected Voat subverses.
As the ecosystem of socio-technical platforms diversifies, it becomes critical to not solely focus on the underlying technologies but also consider the structure of how users interact through the technical infrastructure.
\end{abstract}

\section{Introduction}

Online social spaces are vulnerable to centralized authorities making decisions that negatively affect the community. In 2022, the Software Freedom Conservancy recommended that all developers migrate their projects away from GitHub \cite{intro_sfconservancy}, after Microsoft bought the software development collaboration platform and used open source projects as training data for their commercial CoPilot software, in violation of open source licenses and community standards. The same year, users and advertisers departed Twitter after its purchase by Elon Musk and subsequent changes in community policy and staffing, including firing content moderators \cite{intro_kupferschmidt2022musk} and reinstating a number of accounts banned for violating the platform’s hateful content and harassment policies \cite{intro_musk_impact}. Reddit moderators have historically engaged in blackouts to protest administrative policies \cite{intro_mentions_reddit_blackout}, and these trends are ongoing; in June, 2023, Reddit announced plans to begin charging for API access, sparking warnings from scientists \cite{intro_reddit_davidson2023social}, outrage among users, and a protest across nearly 9,000 subreddits, 
the long-term effects of which remain to be seen. As users express dissatisfaction with platform administrators, they have sought alternative platforms without centralized control, leading to the rapid growth of ``federated" platforms like Mastodon \cite{intro_mastodon_zia2023flocking} and Bluesky\footnote{Bluesky is still in beta, and while the protocol is federated, only one instance exists at the time of writing.}. Alternatively, other users have promoted self-hosted platforms, such as independently operated git servers, or peer-to-peer hosting solutions such as the Interplanetary File System (IPFS) or web-torrent video hosting software PeerTube. Some deplatformed users have also responded by creating close facsimiles of existing centralized platforms with extremely permissive content-policies, frequently called ``alt-tech" platforms \cite{intro_alttech_donovan2019parallel}.

What exactly is ``centralization" in an online social network? Does it describe ownership of the platform? Its technical infrastructure? The creation and enforcement of community norms? The distribution of activity and reach of content producers? Centralization has long been ill-defined by academics \cite{freeman_centrality_1978}, and ``decentralization" joins as a widely-used but contextually redefined term today \cite{di_bona_decentralized_2022}. Of particular interest to us is a notion of group social influence: How much does one community impact information flow across a platform? For example, how independent are subreddits on Reddit, and how closely interlinked are instances on Mastodon, the nascent ``decentralized Twitter alternative?" Our goal is to measure the influence of a platform's sub-communities on their peers by looking at the structure of socio-technical networks, providing a mesoscale metric to quantify centralization at an inter-group level.

One common approach to measuring community-level centralization is through community size-distribution. If a small oligarchy of Mastodon instances dwarf the population sizes of their peers, then one could presume that the platform is centralized around these instances. Indeed, several prior studies on Mastodon use community size disparity as a starting point, or presuppose that the largest instances are the most significant and focus their study on the largest communities \cite{raman_challenges_2019,zignani_follow_nodate,zignani_footprints_2019,la_cava_understanding_2021}. While the community size distribution is related to centralization, assuming they are the same precludes the possibility that a collection of many smaller instances may be more influential than the few largest, or that the largest instances may not represent the platform as a whole.

We reject the assertion that the largest communities must be the most significant, or that their size alone implies centralization, on the grounds that community size does not correlate with the number of cross-community links in observed real-world networks. In fact, our results show multiple platforms where the largest communities are \textit{not} well integrated with the platform as a whole (discussed in \cref{sec:comparison_to_degree_distribution}, especially \cref{fig:voat_render}), allowing a more decentralized network of communities to exist outside of the largest groups. Under this view, the largest communities would be the most significant only when they also act as important information bottlenecks for the entire system.

To illustrate this discrepancy, consider \cref{fig:centralization}. In \cref{fig:centralization-pl} the largest community serves as a central hub, connecting several smaller communities together through shared membership. In \cref{fig:centralization-er}, community size is normally distributed, and no community has a pivotal role as a bridge between its peers. Community size-distribution and graph-centric metrics like betweenness-centrality would agree that the former network is centralized, while the latter is decentralized. However, \cref{fig:centralization-mixed} presents a more complex scenario: the community size distribution is highly unbalanced, but the largest community has no impact on the remainder of the network. Adding one edge to create a complete graph would grant the largest community a high betweenness-centrality because of its pivotal role in connecting so many users to the rest of the graph; but this does not match our intuition that the largest community has a small role in the rest of the network.

We propose a definition of centralization meant to capture the alignment between rankings of community size and information bottlenecks. To do so, we combine theoretical ideas from graph theory on bottlenecks and applied concepts from network science about network resilience. Our metric then measures how removing a community would impact users within remaining communities, based on the number of ``bridges" between communities. We study a variety of real and simulated networks with this method to examine platform behavior under a range of conditions, and we compare our metric to existing measurements of centralization and network ``bottlenecks." Finally, we discuss how this work contributes to broader discussions of centralization online, and how techniques like ours can be extended with richer interaction data.

\section{Prior Work}

In the context of online platforms, centralization is sometimes defined in terms of decision-making power, or who has the authority to make what kinds of decisions about the use of the platform. This definition can be traced to Elinor Ostrom's work on Institutional Analysis and Development \cite{ostrom_background_2011}, which describes ``layers" of decisions, from operational rules (elementary actions any user can perform), to collective rules (the context in which users operate and interact, such as the Twitter feed or the Amazon marketplace), to constitutional rules (the ``meta" rules through which the system changes itself). Modern research on platform design often assesses who has decision-making power, and what levers of change are available to different categories of participants \cite{kraut2012building,frey_this_2019}.

While qualitative studies examine power structures through analyzing governance and rule sets \cite{schneider_modular_2021,fan_digital_2020}, network science infers structure through the observed interactions between humans \cite{freeman_centrality_1978,nadini_mapping_2021}.
We quantify centralization using attributes that fall into three categories: vertex-level attributes, cluster-level attributes, and graph-level attributes. Vertex-level attributes like betweenness centrality \cite{freeman_centrality_1978} or eigenvector centrality \cite{bonacich1987power} measure the prominence of a particular node in terms of how well it is connected to its peers, or how many paths flow through the node. Cluster-level attributes describe groups of vertices, such as the size of the population that contains a particular attribute, or the assortativity describing how likely vertices with a particular attribute are to be connected to one another. Graph-level attributes describe aspects that span the entire network, including diameter, density, and graph conductance \cite{cheeger}. Quantifiability should not be conflated with objectivity; the modeling choice of what entities are included as vertices and what relationships are represented as edges or attributes presupposes what can be considered influential or centralized \cite{butts2009revisiting}.

Another thread of research tries to join the social theory of centralization and graph theoretical metrics. \textcite{fan_digital_2020} distinguish between the technical underpinnings of a network and its social layers, focusing on community-run moderation in infrastructure-centralized (Slack, Discord) and self-hosted (Minecraft) services. Prior Mastodon research also bridges this gap, including both geographic and data-center distribution of instances \cite{raman_challenges_2019}, important for understanding resiliency to disruption or power-outage. This approach aligns with notions of network robustness where centralization can be measured by how a network breaks down under targeted pruning of central nodes \cite{barabasi_attack_tolerance}. Other studies on Mastodon also integrate its social interaction graph \cite{la_cava_understanding_2021}, important for understanding the influence of sub-communities and their administrators on discourse. Studies on the social structure of Mastodon primarily focus on individual-centralization, such as a ``border-index" of what fraction of a user's neighbors are on a foreign instance \cite{zignani_footprints_2019} and whether some users serve as critical bridges for information flow between instances \cite{la_cava_information_2022}, or community-centralization, such as how clustering coefficients differ between communities (instances) \cite{zignani_follow_nodate}. Our work intends to add to these options, by considering both a community-level centralization metric of how much influence one community has on the broader platform, and a graph-level centralization score of how quickly a network deteriorates as its largest communities are removed, indicating how much it tends towards monopoly or oligopoly.

Finally, authors like \textcite{agre_p2p_2003} push back against a Boolean or spectrum from ``centralized'' to ``decentralized,'' and view platforms as layers of centralization and decentralization. Their primary example is eBay, a centrally-controlled marketplace operated by one company, but containing a decentralized network of buyers and sellers. Other studies have built off of this idea, proposing the creation of decentralized ``digital juries" on infrastructure-centralized platforms \cite{fan_digital_2020}, or creating democratic governance bots with flexible bylaws overlayed on centrally-hosted chat services like Slack \cite{zhang_policykit_2020}.
We pose that this layered framework also applies to thinking about complex networks like \cref{fig:centralization-mixed}, which have both centralized and decentralized components that should be considered independently.

\section{Methods and Materials}

We introduce our metric in \cref{sec:disruption_curves}, and two data sets: five real world networks that encompass a breadth of configurations (\cref{sec:real_data}), and a set of common synthetic networks for reference (\cref{sec:disruption_toy}).

\subsection{Measuring centralization: disruption curves} \label{sec:disruption_curves}

Prior studies on centralization of social networks often focus on graph-level attributes such as detecting components, the size of the giant component, modularity, density, degree distribution \cite{ao2022decentralized}. Others may use ``bottleneck" metrics like graph conductance \cite{cheeger} to identify bridges and key clusters. These metrics are most appealing in unipartite settings where the structure of the network is not prescribed. However, we focus on bipartite graphs where communities are well defined, such as subreddits, Mastodon instances, or newsgroups. In these contexts, we are not attempting to infer the number or boundaries of communities, but to measure how influential the known communities are on their neighbors. The size distribution of communities tells us how large a subgroup is, but does not capture the overlap between communities. A graph-wide modularity score describes how well-partitioned the graph is into clusters, and so approximates how insular communities are, but cannot provide more nuance as to whether the largest communities are more integrated than smaller ones, whether small communities are well connected to larger peers but not to each other, or other topological features.

We propose that the influence of a community should be measured in terms of how users outside the community would be impacted by its absence. In other words, a community's influence should be proportional not to its size, but to the number of bridges between it and other communities. Or, in graph theoretic terms, what fraction of edges would be cut by removing a community, not counting users that do not participate outside the community. More succinctly, ``what percentage of edges from surviving vertices would be cut by removing a community?"

We measure disruption cumulatively, rather than discretely per-community. This allows us to answer questions like ``how influential are the largest three communities on the rest of the platform?" Since ``oligarchies" of large and densely interconnected communities may be common, a cumulative metric is more useful than measuring the influence of a single community on the rest of the oligarchy.

Formally, we define a set of communities that are being cut, $A$, with associated edges $|A|$. Each user has a set of edges to one or more communities. If users \textit{only} have edges to communities in $A$, then the user is removed along with $A$. Surviving users with an edge to at least one remaining community are denoted $S$, with total edges $|S|$, and edges to cut communities in $A$ denoted $\partial S$. The disruption curve is calculated as $\partial S/|S|$. This notation was chosen for its similarity to the Cheeger number \cite{cheeger}, stressing how our metric measures the alignment of community size and information bottleneck. We additionally outline the algorithm as pseudocode in supplemental \cref{sec:psuedocode}.

Our disruption curve metric is intended for bipartite networks, where communities are clearly distinguishable and users can participate in multiple communities. However, some consideration is also given to applying our metric to unipartite settings in \cref{sec:unipartite}.

We plot disruption similarly to a cumulative disruption function (CDF), where the x-axis represents the number of communities removed, cumulatively ordered by degree, and the y-axis represents the fraction of edges from surviving users that have been cut. In other words, the x-axis is the size of $A$ as a fraction of all communities in the graph, and the y-axis is $\partial S/|S|$, where both the numerator and denominator are dependent on $|A|$.

While disruption curves offer insight into the role of the largest communities on a platform, some readers may desire a scalar summary statistic to describe how ``centralized" a platform is under our metric. For these scenarios we recommend calculating the area under the curve, as shown in \cref{fig:toy_networks_auc,fig:real_networks_auc}. The Disruption AUC (DAUC) does not indicate how much any particular community influences its peers, but summarizes whether a network is prone to disruption if its largest communities are removed. Methodological choices for calculating the DAUC are discussed in \cref{sec:auc_explanation}.

\begin{table*}[htbp]
    \centering
    \begin{tabular}{l|l|p{6cm}|p{4cm}}
        \textbf{Platform} & \textbf{Community Definition} & \textbf{Edge Definition} & \textbf{Edge weight} \\ \hline
        Mastodon & Mastodon Instances & Between each user and every instance on which they follow users & The number of users followed on an instance \\
        Penumbra & A git server & Between a user (identified by email) and each server on which they have contributed to a repository & The number of repositories committed to on each server\\
        BitChute & BitChute channels & Between each user and every channel they have commented on videos from & The number of comments made \\
        Voat & A Voat ``subverse" & Between each user and subverses they've participated in & Number of comments made in a subverse \\
        Usenet & A Usenet newsgroup & Between each user and every newsgroup they have posted in & The number of posts made
    \end{tabular}
    \caption{Definitions of communities and edges for each platform examined}
    \label{tab:communities_and_edges}
\end{table*}

\subsection{Real-World Network Data} \label{sec:real_data}

We analyze five real-world datasets, each describing online social interactions in bipartite configurations where vertices represent either ``users" or ``communities." We utilize a 2021 scrape of the Mastodon follow graph \cite{zignani_follow_nodate}. Mastodon is a Twitter alternative where users are located on one of thousands of ``instances," which are Twitter-like servers with their own administrators and content policies. However, Mastodon users can follow users on other instances, exchanging content between the two communities, so long as the servers are ``federated" (willing to exchange content). For a second example of a platform with distributed servers, we include the Penumbra of open-source \cite{trujillo2022penumbra}, a data set of independent git servers (not GitHub or GitLab), and users that contribute to repositories on each server. We also include an interaction graph from BitChute \cite{trujillo2022mela}, an alt-tech YouTube alternative, consisting of users and the channels (video uploaders) whose videos they commented on. We utilize a similar scrape of Voat \cite{voat2022}, an alt-tech Reddit alternative active until late 2020, consisting of users and the ``subverses" (subreddits) they commented in. We additionally include an archive of Polish Usenet groups \cite{usenet_polish_archive}, providing a much older but similarly structured platform for comparison. Details on the vertex and edge definitions for each network are included in \cref{tab:communities_and_edges}, and the size of each network is listed in \cref{tab:network_sizes}.

\begin{table}[htbp]
    \centering
    \begin{tabular}{l|l|l|l}
        \textbf{Platform} & \textbf{Comms.} & \textbf{Users} & \textbf{Edges} \\ \hline
        Mastodon & 3,825 & 479,425 & 5,649,762 \\
        Penumbra & 841 & 41,619 & 108,038 \\
        BitChute & 29,686 & 299,735 & 11,549,058 \\
        Voat & 7,515 & 3,624,486 & 16,263,309 \\
        Usenet & 333 & 2,080,335 & 58,133,610
    \end{tabular}
    \caption{Population size of each network in terms of community count, user count, and relationship edge count, before compressing duplicate edges into weighted edges}
    \label{tab:network_sizes}
\end{table}

We selected these platforms because they have clear bipartite user and community representations, their data is readily available, and each platform is small enough to obtain a nearly-complete sample. Sub-sampling a larger platform like Reddit is likely to miss lower-population or lower-activity sub-communities, and we are particularly interest in the interactions between smaller communities. The resulting dataset encompasses a variety of approaches to hosting and community governance, providing a spectrum of ``centralization."

\subsection{Synthetic Network Data} \label{sec:disruption_toy}

To understand disruption curves and contextualize our real-world results, we examine a variety of well understood synthetic network topologies.

First we construct a bipartite star network, as a default example of a network centralized around a single hub. In our example plots, we construct a graph with 150 communities and 3000 users, such that every user has an edge to two communities: the central hub, and one other, assigned uniformly. Removing the hub eliminates 50\% of all edges, and removing any subsequent communities incurs no additional disruption, because all impacted users will have a degree of zero and be pruned from the graph (see \cref{fig:toy_networks_size_comparison}). This graph type is therefore highly centralized but has a decentralized periphery after the removal of the central community, illustrating how different topologies can co-exist in the same network, muddying the definition of ``centralization."

We then test disruption on a variety of bipartite networks with power-law degree distributions. We first adapt the Barab\'{a}si-Albert preferential attachment model to a bipartite setting, initializing a network with 300 empty communities and introducing users that connect to a given community with probability proportional to their size plus one. We also introduce a range of bipartite configuration models: in each, we assign a degree to each community drawn from a power law with a specified $\gamma$ exponent. For each community, we create edges according to degree, connecting the community to users uniformly randomly without replacement. Therefore, we do not control for user degree or assortativity, only for the size of communities. Each of these networks produces a curve that slowly decays towards a diagonal, implying that removing the largest communities has some disproportionate impact, after which removing additional communities has a less pronounced result.

For a contrast from power-law degree distributions, we also adapt the Erd\H{o}s-R\'{e}nyi model to a bipartite setting, by creating vertices for communities and users, then creating all possible edges with a probability $p$ (in our tests, $p=0.05$), while preserving the bipartite constraint. These networks produce a disruption curve with a second derivative near zero, indicating that most communities have near-equal influence on the population, and so removing the largest communities does not have a much larger impact than removing subsequent communities.

Lastly, we adapt the Watts-Strogatz small-world model to a bipartite setting. To accomplish this, we first produce a \textit{unipartite} network with desired neighborhood size ($n=5$) and edge density ($p=0.05$) parameters. Then we apply a clustering algorithm (in our examples we have used weighted community label propagation) to place each user in one community. We create a vertex for each detected community, and replace all user-user edges with user-community edges. This process is described in greater detail in \cref{sec:unipartite} and an example is illustrated in \cref{fig:unipartite}. These networks have the most uniform community size distribution of any we tested, and their disruption curves are similar to those of Erd\H{o}s-R\'{e}nyi networks, with slightly more variability.

\begin{figure*}[htb]
    \centering
    \hfill
    \begin{subfigure}[b]{0.45\textwidth}
        \centering
        \includegraphics[width=\textwidth]{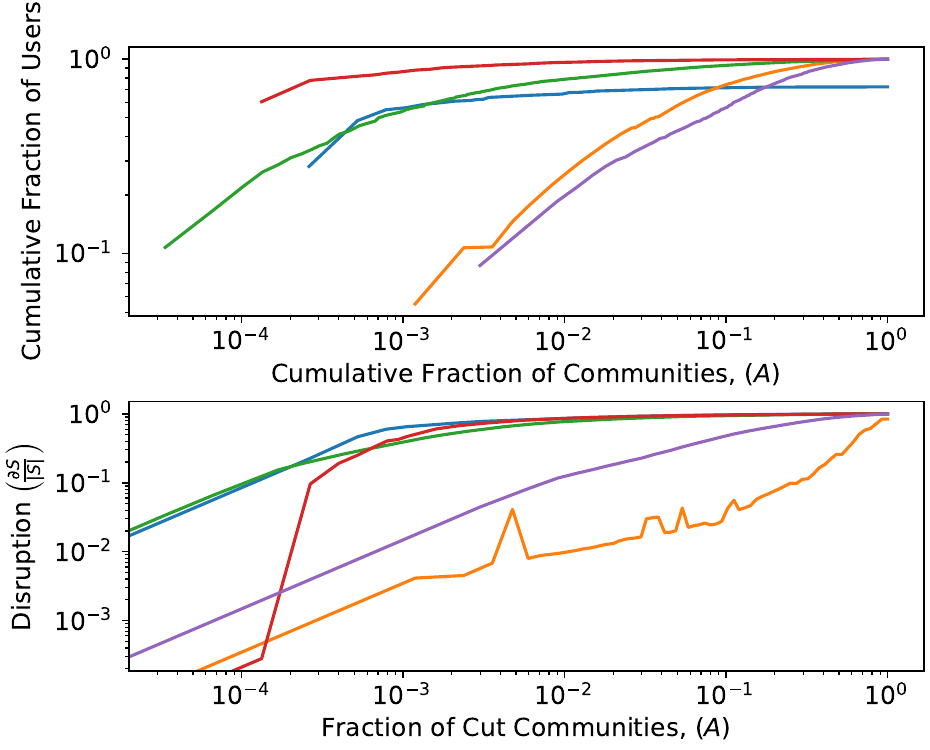}
        \caption{Community size distribution and disruption curves}
        \label{fig:real_networks_size_comparison}
    \end{subfigure} \hfill
    \begin{subfigure}[b]{0.45\textwidth}
        \centering
        \includegraphics[width=\textwidth]{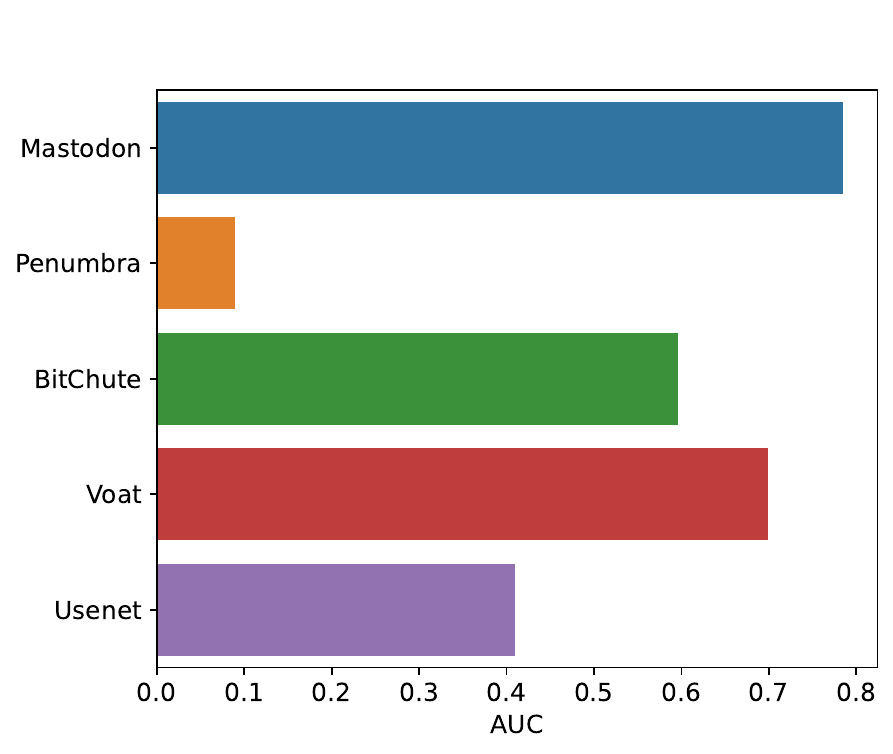}
        \caption{Area under the disruption curve (DAUC)}
        \label{fig:real_networks_auc}
    \end{subfigure}
    \hfill

    \caption{Summary measures of centralization. 
    (a)
    Our measure of community disruption (bottom) does not correlate with the population distribution of communities (top). 
    (b) 
    The area under the disruption curve (DAUC) provides a summary statistic of the disruption curve that reinforces how network structure combined with community size provide greater insight into centralization (measurement details in \cref{sec:auc_explanation}). 
    Here, panel (a) consists of cumulative distribution plots of population and disruption, where the top subplot is a CDF of the platform population as smaller communities are included, and the bottom subplot shows how networks are damaged as more of the largest communities are removed. 
    Each line represents a different network, using the color key from panel b.}
    \label{fig:real_networks_main_figure}
\end{figure*}

For Erd\H{o}s-R\'{e}nyi and configuration model networks, it is possible to create a closed-form solution for cumulative disruption. 
This is not our emphasis, because we are primarily concerned with analyzing real-world networks for which network generating functions are not available, but we detail this formalism in supplemental \cref{sec:analytic_simulations}. 
We also include further descriptions and reference visualizations for bipartite near-star and bipartite power-law networks in \cref{sec:toy_examples}.

\section{Results}

\begin{figure*}[htb]
    \centering
    \hfill
    \begin{subfigure}[b]{0.45\textwidth}
        \centering
        \includegraphics[width=\textwidth]{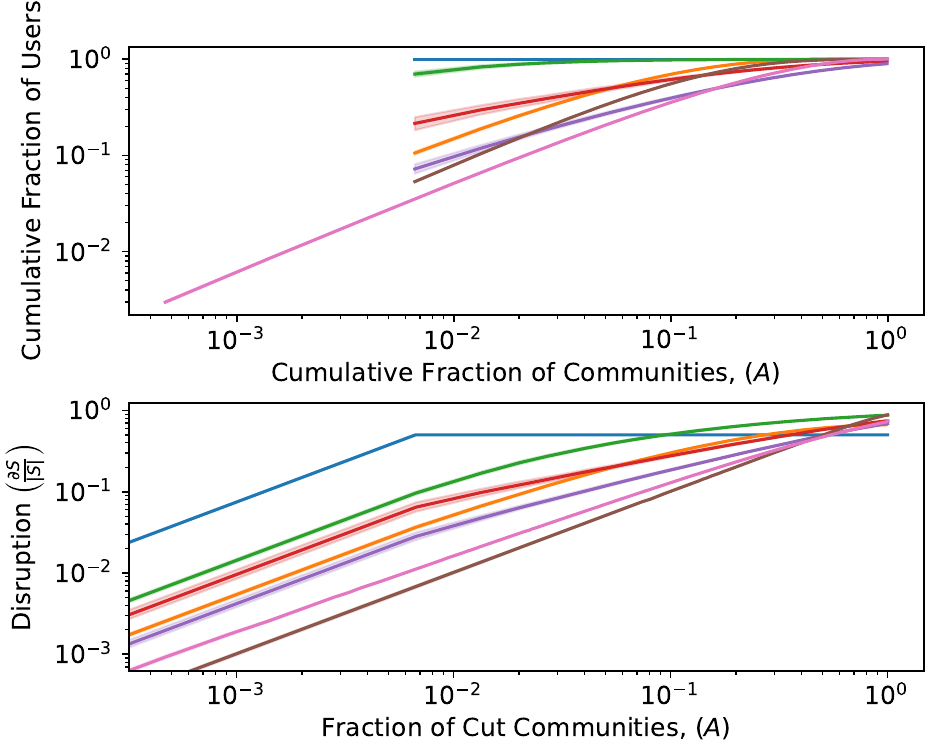}
        \caption{Community size distribution and disruption curves}
        \label{fig:toy_networks_size_comparison}
    \end{subfigure} \hfill
    \begin{subfigure}[b]{0.45\textwidth}
        \centering
        \includegraphics[width=\textwidth]{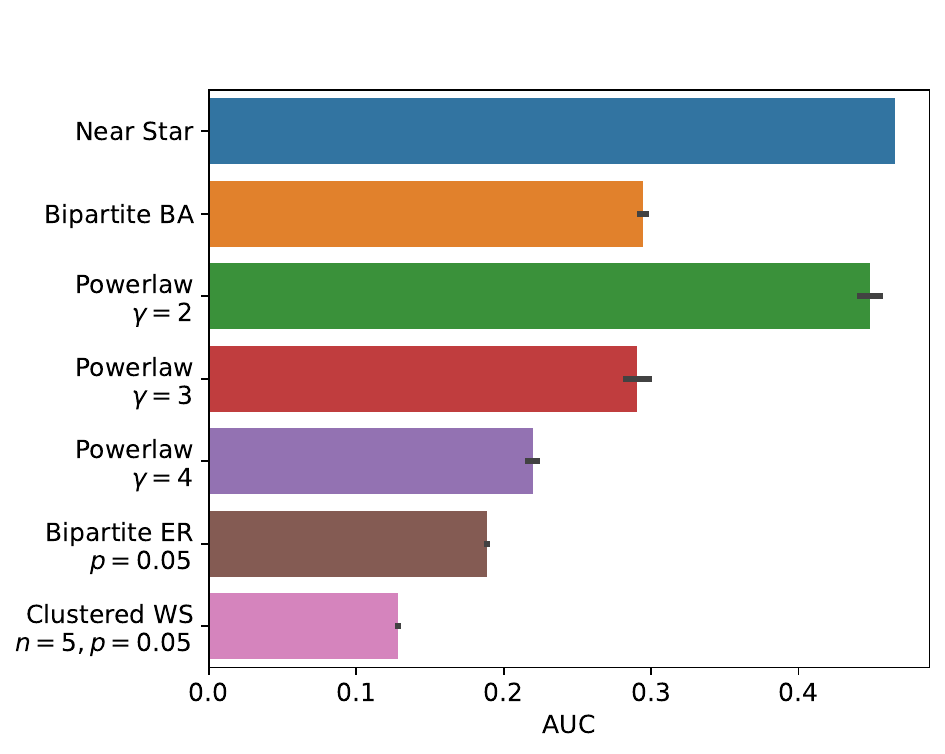}
        \caption{Area under the disruption curve (DAUC)}
        \label{fig:toy_networks_auc}
    \end{subfigure}
    \hfill

    \caption{In simulated networks with a variety of degree distributions, the disruption curves for each network much more closely match the population distribution (\cref{fig:toy_networks_size_comparison}), suggesting that non-degree network attributes such as assortativity play a crucial role in determining centralization. As in \cref{fig:real_networks_main_figure}, the left figure represents cumulative population and disruption as more communities are considered. Each line represents a network sharing the color-key in the right figure. Simulated networks were generated 100 times, and the mean and a 95\% confidence interval are shown in both figures.}
    \label{fig:toy_networks_main_figure}
\end{figure*}

We plot the cumulative population size, disruption curve, and disruption AUC for real-world networks in \cref{fig:real_networks_main_figure}, and plot the same results for synthetic network data in \cref{fig:toy_networks_main_figure}. We first focus on discrepancies between the size distribution and disruption curves for real networks in \cref{sec:comparison_to_degree_distribution}, then return attention to synthetic network data when we examine the role of assortativity in \cref{sec:assortativity}.

\subsection{Comparison to Size Distribution} \label{sec:comparison_to_degree_distribution}

Upon comparing the size distribution and disruption curve in \cref{fig:real_networks_size_comparison}, it is apparent that the community size distribution is insufficient to describe the structure of a network. 
Voat has the most skewed population distribution: almost all users participate in the largest community, yet the network does not experience significant disruption until the largest \textit{three} communities are removed. 
Mastodon and BitChute have the next most skewed size distributions, but there is a large distance between the proportional sizes of their largest communities, and almost identical disruption curves as those communities are removed. 
By population distribution, the Penumbra appears to be more skewed towards its largest git servers than Usenet is towards its largest newsgroups.
This is not mirrored in disruption curves, where Usenet has a consistently higher disruption than the Penumbra.

\begin{figure}[htb]
    \centering
    \includegraphics[width=\linewidth]{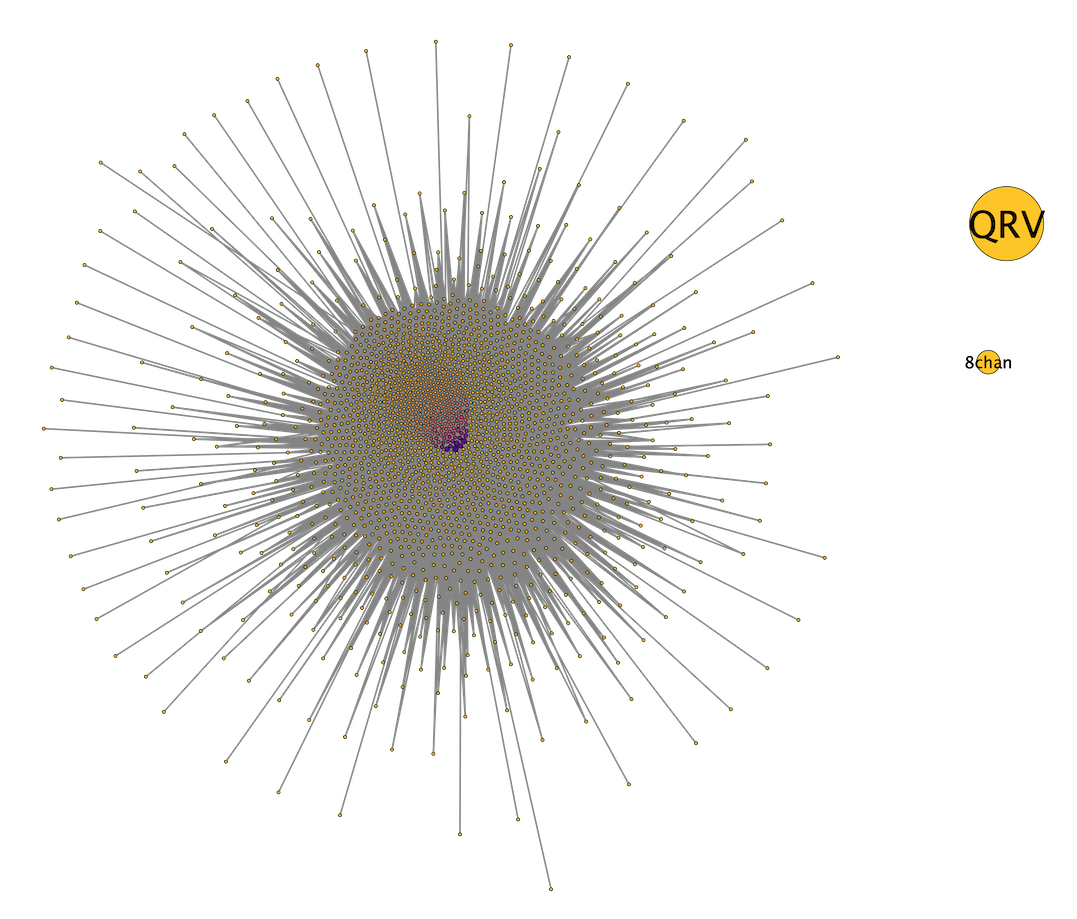}
    \caption{The two largest Voat communities (`QRV' and `8chan') are dramatically larger than their peers, but have almost no overlap in population, making community size a poor proxy for platform-wide influence or centralization. 
    In this network visualization, nodes represent Voat ``subverses," and edges represent at least thirty shared users active in two communities. Node size correlates with user count, and color correlates with strength; i.e. the level of overlap with neighboring communities. The purple communities at the center are default subverses all new users are subscribed to (``news," ``whatever," etc), surrounding pink and orange communities are popular with lots of user overlap. The largest two communities, ``QRV" and ``8chan," have almost no user overlap with other communities and are rendered to the right.}
    \label{fig:voat_render}
\end{figure}

To explain these discrepancies, we examine each network in greater detail. Voat was a Reddit-like platform where users commented and posted in one or more ``subverses.'' 
While users chose to subscribe from among 7515 public subverses, new accounts were automatically subscribed to a set of 27 subverses by default. 
This ``default subscription" has no parallel on other platforms we examined.
Since these default subverses have an automatic population, they are more likely to receive engagement than subverses that must be discovered according to a user's area of interest, and we may expect them to be densely connected with most users on the platform. However, the largest two subverses on Voat by number of unique users were \textit{not} default subverses; \texttt{v/QRV} was a QAnon conspiracy group, and \texttt{v/8chan} was a right-wing news and discussion forum whose name references the white supremacist imageboard 8chan (now ``8kun"). Both subverses were highly insular, with little population overlap with the rest of the platform, as illustrated in \cref{fig:voat_render}. Therefore, it is only when we remove the \textit{third-largest} subverse, \texttt{v/news}, that we see a large impact on remaining users on the site.

The Penumbra of open-source represents software development on git servers outside of GitHub and the primary GitLab instance. Each community represents a git servers with one or more public repositories, and edges indicate that a user (identified by email address) contributed to a repository on a server. Servers are often created per-organization; for example, the Debian Linux distribution hosts their own GitLab server at \url{salsa.debian.org}, and the University of Vermont hosts a GitLab server at \url{gitlab.uvm.edu}. Users often contribute to multiple repositories on a single server, but connections \textit{between} servers are extremely sparse. This sparsity is responsible for the ``spikes" in the Penumbra's disruption curve; removing a git server may sever an edge to some users, and removing a second, related server may prune all remaining edges to those same users. When the cross-server collaborative users are removed, the impact on the remaining less-collaborative community decreases. In all other networks enough users have a sufficiently high cross-community degree that disruption only increases as communities are removed.

Rather than examining the cumulative community size distribution, one could instead examine the size of the giant component of each network. The giant component will shrink as communities are cumulatively removed, providing another means of examining the influence of large communities. This is conceptually similar to our disruption metric, with a critical difference: inclusion of a community in the giant component is Boolean. If a community is completely insulated from the giant component of a network, then its removal will have no effect on the component size; otherwise the component will shrink by the size of the users removed with the community. Our metric provides greater flexibility for measuring how well integrated a community is among its peers. Measurement of the giant component size is discussed further in supplementary \cref{sec:giant_components}.

In summary, the community size distribution cannot adequately describe the topology of a platform because it does not account for features including assortativity or sparsity.

\subsection{Assortativity and Centralization} \label{sec:assortativity}

\begin{figure}[hbt]
    \centering
    \includegraphics[width=\linewidth]{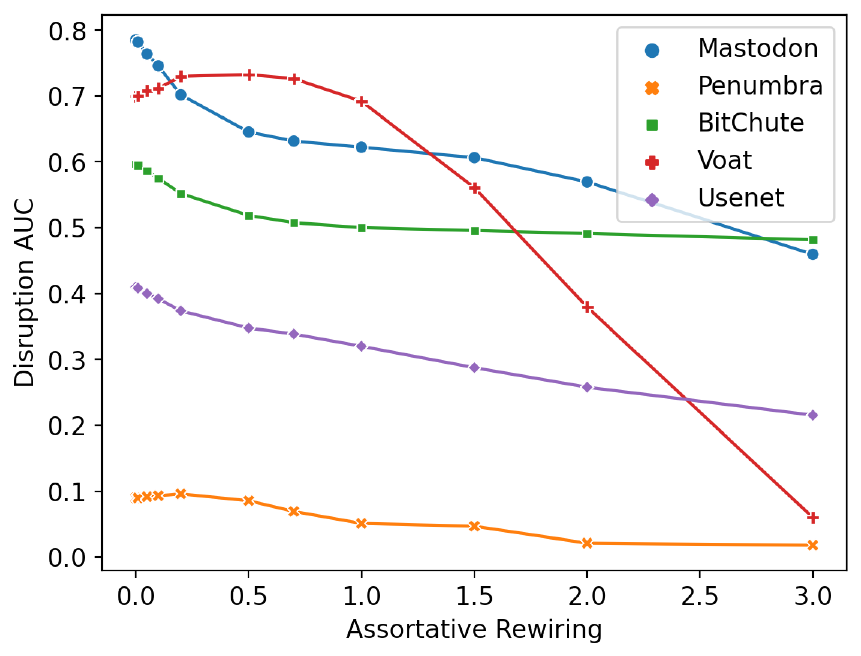}
    \caption{Increasing user-community degree assortativity through edge-rewiring increases the influence of the largest communities in highly insular (Voat) or sparse settings (Penumbra), but decreases disruption in all networks as increased rewirings eliminate cross-community edges and yield insular and sparse networks. Y-axis represents disruption AUC (see \cref{fig:real_networks_auc}), so that the slope shows change in disruption AUC as networks are rewired to increase user-community degree assortativity.}
    \label{fig:assortativity_auc}
\end{figure}

We expect that degree assortativity (or degree-degree correlation) plays a significant role in the differences between observed community disruption (\cref{fig:real_networks_size_comparison}) and network behavior under controlled degree distributions (\cref{fig:toy_networks_size_comparison}). In a purely random setting, users are likely to have edges to multiple large communities, because most edge stubs in a configuration model come from high-degree communities. In real social settings, the content of communities may inhibit assortativity, as in Voat, where the largest two communities are highly insular (see \cref{fig:voat_render}), creating a large disparity between the community size distribution and disruption metric. 

To explore this hypothesis, we randomly rewired each social network to increase assortativity. We select pairs of edges uniformly without replacement, and swap the communities of the edges if it would increase user-community degree assortativity. We continue this process until we have rewired a desired percentage of edges; if we exhaust the edge supply before finding sufficient valid swaps, we re-shuffle the edge list and continue drawing. 
For each rewired network we calculate its disruption and the area under the disruption curve, as in \cref{fig:real_networks_auc}, and plot the change in AUC during rewiring in \cref{fig:assortativity_auc}.

This experiment is useful in distinguishing the idea of network centralization from classic ideas of monopoly. These are two different, but related, problems that are easy to confuse when focusing solely on summary statistics like community size distributions. When a network consists of disconnected communities, it is decentralized under the disruption metric regardless of the size distribution of these communities. This conclusion follows from our definition of centralization since removing a community in a sparse (or disconnected) network, has little (or no) impact on other communities. This rewiring experiment highlights this logic: As networks get rewired to increase correlations, we increase the likelihood of having all the activity of a user focused on a single community and therefore progressively disconnect the community and decentralize the network. The only exception is Voat, whose initial state contains large disconnected communities that can get coupled to the rest of the network by rewiring, before being re-disconnected as we rewire more and more. Small correlations in large networks can therefore increase centralization, since large communities can broker more bridges when they contain well-connected users; while strong correlations in smaller networks can decrease centralization by focusing user activity on single communities.

We analyze other assortativity metrics in additional detail in the supplemental material, \cref{sec:supplemental_assortativity}, as well as a comparison to the Cheeger number ``bottleneck" metric \cite{cheeger}, in \cref{sec:cheeger}. We also confirm our intuition about the role of weak correlations with mathematical analysis of random infinite networks in supplemental \cref{sec:analytic_simulations}.

\section{Conclusion and Future Work}

We have added to the wealth of centralization metrics by proposing a mesoscale measurement that indicates how much influence one sub-community has over a broader network, by accounting for how many edges to remaining users would be severed if a community were removed. This metric allows us to differentiate between networks with a substantial community size-imbalance, and networks where the largest communities play a core structural role in their smaller peers. We extend our metric to create a graph-level measurement that indicates how ``oligopic" a network is, or how well-integrated its largest communities are with the population at large.

We have utilized our disruption metric to examine a range of real-world social networks, comparing their network topology, distribution of community sizes, and the influence of those communities. We find that some platforms, like Voat, are much less centralized than their skewed community-size distribution would suggest, while others, like Usenet and the Penumbra of Open-Source, have similar size distributions and widely divergent disruption curves. Mastodon, while vocally supportive of decentralization, has a disruption curve mostly characterized by the skewed population distribution of its sub-communities and is therefore relatively centralized.

Using simulated networks with a range of degree distributions, and rewiring techniques to adjust assortativity, we have begun to explore the interplay between community size, structure, and community-level centralization. However, we limited ourselves to traditional network generative models like Erd\H{o}s-R\'{e}nyi and power-law configuration model networks. Future research could directly simulate networks with chimeric centralization which combine decentralized and centralized components to more realistically represent the diversity observed in social networks.

Our network representations are oversimplified in that we assume that each edge on a network represents a path of information flow. However, one user following another represents \textit{potential} information flow; a bridge between two communities is only realized if the following user is online and chooses to propagate information from the edge to their own followers and instance.

More thorough research should examine how many potential bridges are utilized by, for example, monitoring the number of ``boosts" (Mastodon's equivalent to ``retweets") across instance boundaries on Mastodon. Observed information spread, and examining the reception of cross-pollinated ideas in non-originating communities, would provide much greater insight into how multi-community platforms function in practice.

\subsection*{Competing Interest Declaration}

The authors declare no competing interest.

\subsection*{Data, Documentation, and Code Availability}

Real-world networks used in this study are available from \cite{zignani_follow_nodate}, and \cite{trujillo2022penumbra,trujillo2022mela,voat2022,usenet_polish_archive}. Code used in this research, including scripts for generating our synthetic networks, will be submitted to a public archive before publication.

\subsection*{Acknowledgements}

All authors were supported by Google Open Source under the Open-Source Complex Ecosystems And Networks (OCEAN) project. 
Any opinions, findings, and conclusions or recommendations expressed in this material are those of the authors and do not necessarily reflect the views of Google Open Source.

\printbibliography

\beginsupplement
\begin{refsection}

\section{Pseudocode Example of Cumulative Disruption Algorithm} \label{sec:psuedocode}

For readers seeking a succinct code-like description of our cumulative disruption curve algorithm, we have included \cref{lst:psuedocode}.

\begin{lstlisting}[label=lst:psuedocode, language=Python, caption=Pseudocode for disruption algorithm]
disruption = []
for c in communities:
    remaining = 0
    original = 0
    removeCommunity(c)
    for user in users:
        if degree(user) > 0:
            remaining += degree(user)
            original += originalDegree(user)
    disruption += [1 - (remaining / original)]
\end{lstlisting}

Note that when calculating disruption on large networks, it is much more efficient to cache the size of the smallest community that each user participates in. We can then sort all users by the order in which they will be removed, and avoid computationally expensive references to a graph or adjacency matrix for each removal-step in the algorithm.

\section{Applications to Unipartite Networks} \label{sec:unipartite}

Our influence metric is intended for settings with clearly defined communities. For example, participation in subreddits, membership on a Mastodon server, or committing to a software code repository, all discretely identify users as members of those explicitly-bounded groups. However, network data is often presented in a unipartite configuration such as users following other users. If it is still desirable to delineate communities and measure their influence in these settings, then they can be converted into compatible bipartite networks using the following procedure:

\begin{enumerate}
    \item Apply a context-appropriate community detection algorithm to label each user as belonging to one community

    \item Create a vertex for each community

    \item Replace all user-user edges with user-to-community edges, where the edge weight is equal to the number of unipartite edges each user had to other nodes in that community

    \item Apply our influence metric to the resulting bipartite graph
\end{enumerate}

\begin{figure}[hbtp]
    \centering
    \includegraphics[width=\linewidth]{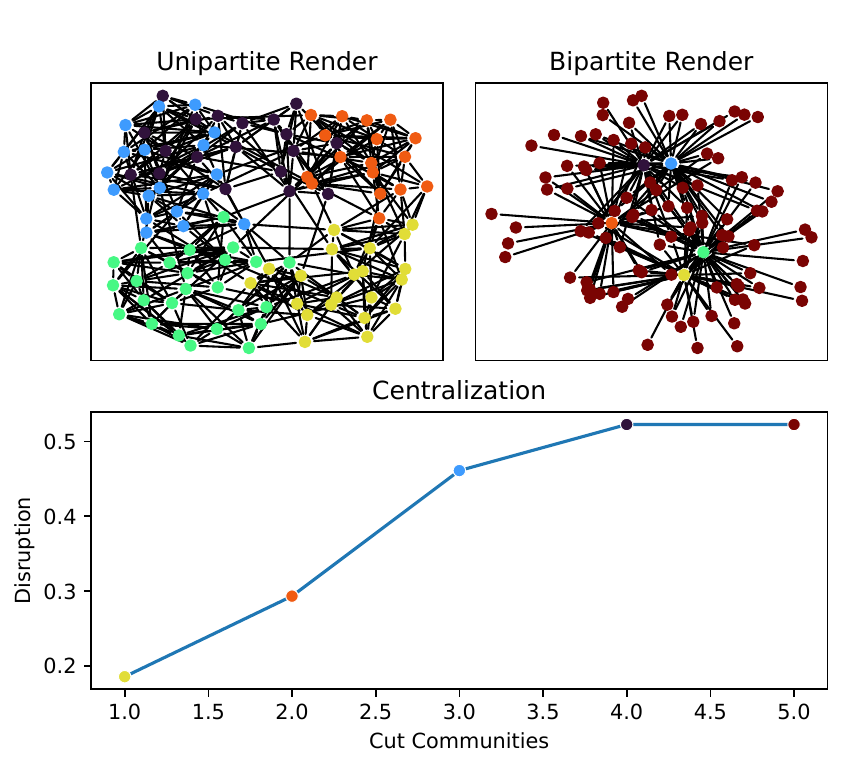}
    \caption{Example of applying our disruption metric to unipartite graphs by
    detecting communities on a unipartite small-world network (top-left), converting labeled communities into a bipartite representation (top-right), and running our influence metric on the bipartite graph (bottom)}
    \label{fig:unipartite}
\end{figure}

An example of this procedure is illustrated in \cref{fig:unipartite}, using a unipartite Watts-Strogatz small-world network (100 nodes, 5 neighbors, rewiring probability of 5\%), and label-propagation for community detection. The unipartite graph is shown in the top-left with community labels visualized with color. It is converted to a bipartite representation shown in the upper-right, and the effect of removing each community is illustrated in the bottom frame.

\section{Calculating the Area Under the Disruption Curve} \label{sec:auc_explanation}

For \cref{fig:real_networks_auc,fig:toy_networks_auc,fig:assortativity_auc} we use the area under the disruption curve as a single-variable summary of how centralized a network is around its largest communities. To calculate the AUC, we use a trapezoidal approximation in logarithmic space.

We chose a trapezoidal approximation to calculate the area even with limited sample points from real-world networks. Integration is possible for purely analytic disruption curve simulations as in \cref{sec:analytic_simulations}, but this is not feasible for our non-Erd\H{o}s-R\'{e}nyi networks, so we use a trapezoidal approximation for all synthetic networks for consistency.

We measure the AUC in logarithmic space, because measuring in linear space would heavily weight the influence of the smallest communities that are removed last, and our primary interest is in examining the influence of the largest communities on the broader population. 

\section{Synthetic Network Topology Details} \label{sec:toy_examples}

We measure centralization on a variety of synthetic networks introduced in \cref{sec:disruption_toy}. In this section, we include further description and visualization of the synthetic networks used.

\begin{figure}[htb]
\centering
\begin{tikzpicture}
\Vertex[x=0,y=0]{H}; 

\Vertex[x=-1,y=-1,shape=rectangle]{L1};
\Vertex[x=-1,y=1,shape=rectangle]{L2};
\Vertex[x=1,y=-1,shape=rectangle]{L3};
\Vertex[x=1,y=1,shape=rectangle]{L4};

\Vertex[x=-2,y=-2]{C1};
\Vertex[x=-2,y=2]{C2};
\Vertex[x=2,y=-2]{C3};
\Vertex[x=2,y=2]{C4};

\Edge(H)(L1);
\Edge(H)(L2);
\Edge(H)(L3);
\Edge(H)(L4);

\Edge(L1)(C1);
\Edge(L2)(C2);
\Edge(L3)(C3);
\Edge(L4)(C4);

\Vertex[x=1.7,y=0.3,shape=rectangle]{L5};
\Vertex[x=0.3,y=1.7,shape=rectangle]{L6};
\Edge(H)(L5);
\Edge(L5)(C4);
\Edge(H)(L6);
\Edge(L6)(C4);

\Vertex[x=-1.7,y=-0.3,shape=rectangle]{L7};
\Vertex[x=-0.3,y=-1.7,shape=rectangle]{L8};
\Edge(H)(L7);
\Edge(L7)(C1);
\Edge(H)(L8);
\Edge(L8)(C1);

\end{tikzpicture}
\caption{Example Bipartite Near-Star. Circles are communities, squares are users. All users are connected to two communities, and a primary hub is connected to all users.}
\label{fig:star}
\end{figure}
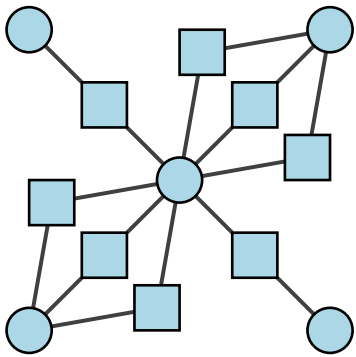

Bipartite Near-Star networks are analogous to a unipartite star network with duplicate edges, but in a bipartite setting. Starting with a unipartite star, replace each edge from the hub to a leaf with a two-path from the hub community to a new ``user" vertex, to the leaf community. Duplicate edges from the unipartite hub to leaves are converted into multiple users that share a community, and serve to break ties when pruning communities for disruption curves. This is illustrated in \cref{fig:star}.

For our ``Powerlaw" networks we follow a bipartite configuration model. We first create vertices representing the desired number of communities and users. We then draw from a powerlaw distribution with an assigned $\gamma$ exponent, and assign the drawn degree to each community. Then, we create a corresponding number of edges, wiring each community to users drawn uniformly at random without replacement. This yields networks where communities follow a powerlaw degree distribution, while users follow a normal degree distribution.

\begin{figure}[htb]
    \centering
    \begin{tikzpicture}[multilayer=3d,scale=0.7]
\SetLayerDistance{-2.5}
\Plane[x=0.0,y=1.0,width=4.5,height=4.0,color=blue!20,layer=3]
\Plane[x=0.0,y=1.0,width=4.5,height=4.0,color=green!20,layer=4]
\Vertices[shape=circle,size=0.2,NoLabel=True]{centralization-figure/powerlaw_vertices.csv}
\Edges[color=black!20]{centralization-figure/powerlaw_edges.csv}
\begin{Layer}[layer=3]
    \node at (0.0,1.0)[below right]{Communities};
\end{Layer}
\begin{Layer}[layer=4]
    \node at (0.0,1.0)[below right]{Users};
\end{Layer}
\end{tikzpicture}
    \caption{An example bipartite powerlaw network, visualized using a ``community" and ``user" layer}
    \label{fig:pl-toy}
\end{figure}
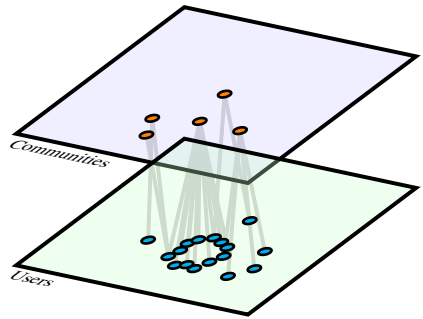

Bipartite community-user networks can be visualized in a flat plane, as in \cref{fig:centralization-pl}, or as a multi-layer graph, as in \cref{fig:pl-toy}. A multi-layer representation may be beneficial for representing inter-community relationships that are not explained by shared users, such as Mastodon federation agreements, or shared moderator staff in two subverses. However, these multiplex relationships were deemed out-of-scope for our current work.

\section{Mathematical Analysis of Disruption in Random Networks} \label{sec:analytic_simulations}

We here calculate the disruption curves for random bipartite networks parameterized by their joint-degree distribution. This approach therefore fixes the distribution $\lbrace g_m \rbrace$ of communities $m$ per user, the distribution $\lbrace p_n \rbrace$ of community size $n$, and the joint-distribution $P_{n,m}$ for the degree of the node and community involved in a random bipartite link. Beyond these constraints, the networks are fully random but allow us to explore the role of heterogeneous connectivity at the user and community level as well as the impact of correlations between both levels.

We wish to calculate the disruption $D(n)$ involved when removing communities of size $n'<n$ in these random networks. By definition of the bipartite network, we know that $np_n$ edges are removed when removing communities of size $n$. Once again, we define disruption as the fraction of \textit{remaining} edges disrupted by communities of size $n$ during the pruning process. It is thus given by the number of edges that belong to communities of size $n$ minus the fraction $u_n$ of those that are the sole edge of the corresponding users (since these users are removed in the pruning) divided by the number of edges belonging to communities of size equal or smaller than $n$ minus the $u_nnp_n$ users removed. We write:

\vspace{2em}
\begin{equation}
    D(n) = \frac{
            \eqnmarkbox[NavyBlue]{bigedges}{np_n}
            -
            \eqnmarkbox[OliveGreen]{prunededges}{u_nnp_n}
        }{
            \eqnmarkbox[WildStrawberry]{remainingedges}{\sum_{n'\leq n}n'p_{n'}}
            -
            \eqnmarkbox[OliveGreen]{prunededges2}{u_nnp_n}
        } \; .
\end{equation}
\annotate[yshift=1em]{above,left}{bigedges}{Edges to comms. of size n}
\annotate[yshift=1em]{above,right}{prunededges}{Edges to removed users}
\annotate[yshift=-0.5em]{below}{remainingedges}{Edges to comms. n or smaller}
\vspace{2em}

The quantity $u_n$ can also be defined as the probability that a random user of a community of size $n$ has no community smaller than $n$. It can therefore be calculated like so:

\vspace{1em}
\begin{equation}
    u_n = \mathlarger{\sum}_m 
        \eqnmarkbox[NavyBlue]{users_in_n_with_m}{\frac{P_{n,m}}{\sum_{m'}P_{n,m'}}}
        \left(
            \eqnmarkbox[OliveGreen]{users_with_m_larger_than_n}{\frac{\sum_{n'\geq n} P_{n',m}}{\sum_{n'}P_{n',m}}}
        \right)^{m-1} \; .
    \label{eq:un}
\end{equation}
\annotate[yshift=1em]{above,right}{users_in_n_with_m}{Fraction of users in comm. \\ \sffamily \footnotesize size n that have m edges}
\annotate[yshift=-0.5em]{below,left}{users_with_m_larger_than_n}{Fraction of users with m edges\\ \sffamily \footnotesize in comms. larger than size n}
\vspace{2.5em}

In the previous equation, we sum over every possible type of node in a community of size $n$, which will have a number of \textit{other} communities $m-1$ proportional to $P_{n,m}$, and ask for all of these communities to be larger or equal to $n$, which will be proportional to the sum of $P_{n',m}$ over all $n'$ larger or equal to $n$. Normalizing the probabilities appropriately yields Eq. (\ref{eq:un}) as written.

Note that these equations assume that edges are unweighted, and that there are no duplicate edges, which is what we expect from an infinite random simple graph. In our real-world data sets there are often duplicate edges (for example, one user following several different users on a Mastodon instance), which we compress to weighted edges for convenience.

Despite this difference between the analytical expression and real socio-technical networks, the analysis of random infinite graphs can be useful to test how disruption is impacted by simple network statistics such as degree distributions or correlations in the joint community-user degree matrix $P_{n,m}$. 

In a simple experiment, we create a random Erd\H{o}s-R\'{e}nyi-like bipartite network and correlated equivalent networks with the same degree distributions and variable community-user degree matrices $P_{n,m}$. The random network has a simple $P^{\textrm{rand}}_{n,m} \propto np_n mg_m$ (normalized) which we can modify manually. To do so, we calculate the maximally correlated $P^{\textrm{max}}_{n,m}$ by assigning users with highest degrees $m_{\textrm{max}}$ to the largest communities available before doing the same to users with the next higher degree and so on all the way down. We can do the same to calculate $P^{\textrm{min}}_{n,m}$ by assigning users with the lowest degree to the largest communities and working our way up in the user degree distribution. We can then create arbitrary community-user degree matrix $P_{n,m}$ by interpolating between linearly with $(1-\rho) P^{\textrm{rand}}_{n,m} + \rho P^{\textrm{max}}_{n,m}$ or $(1-\rho) P^{\textrm{rand}}_{n,m} + \rho P^{\textrm{min}}_{n,m}$.

\begin{figure}[ht!]
    \centering
    \includegraphics[width=0.9\linewidth]{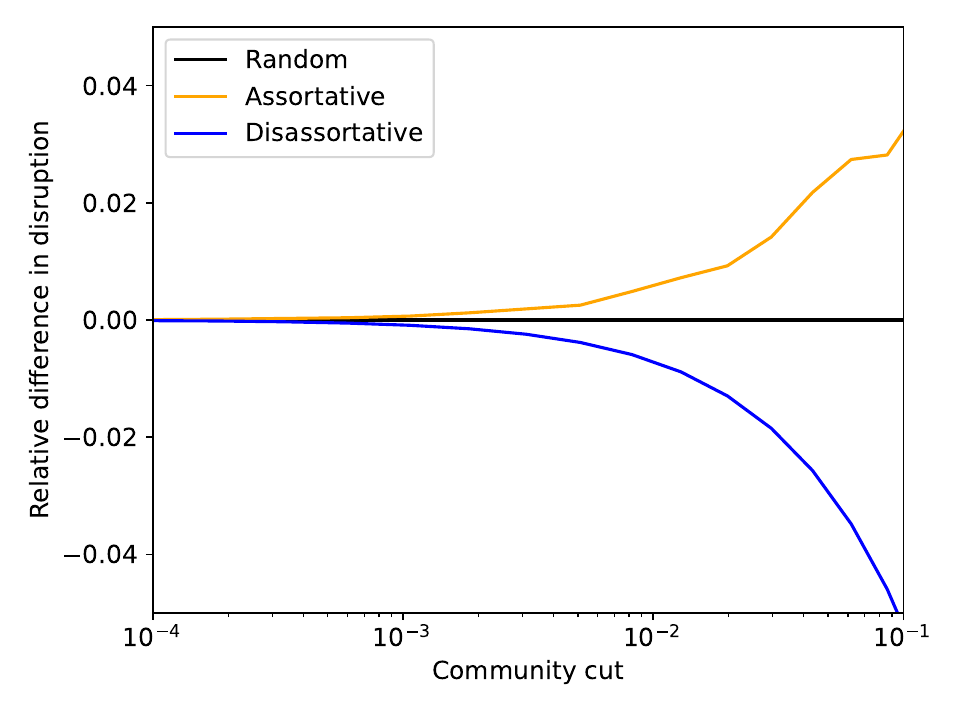}
    \caption{Random bipartite networks with varying user-community degree correlations. We start with a random bipartite network where the ratio of users to communities is 30 and a binomial distribution of communities per user with average 1.2. We create two counterfactuals where degree correlations are at 30\% of the maximally assortative network and of the maximally disassortative network. We show the relative difference in disruption caused by correlations (disruption of correlated network divided by disruption of random network minus 1).}
    \label{fig:assortivity_random_networks}
\end{figure}

Our results are shown in \cref{fig:assortivity_random_networks}. We find that positive user-community degree correlations increase disruption and therefore \textit{centralizes} the resulting socio-technical network. Conversely, negative correlations decreases correlations and \textit{decentralizes} the network. That being said, the relative effect of correlations is relatively small as the networks are still otherwise completely random.

\section{Further Analysis of Assortativity} \label{sec:supplemental_assortativity}

There are multiple interpretations of degree assortativity in a bipartite setting. The linear correlation between user degrees and community degrees measures whether high-degree users are likely to be connected to high-degree communities. In our network definitions edges represent activity, like follow relationships or participation in conversations, so this measures whether active users are likely to be connected to communities with lots of activity. However, a second metric of interest is whether large communities are likely to be connected to other large communities, or in other words, the  assortativity of a unipartite-projected community-community graph. This can also be broken into two sub-cases: assortativity of community size (do communities with many users share users with other high-population communities), and assortativity of degree (do communities with lots of activity share users with other high-activity communities).

These three notions of assortativity are not independent; we might expect that users with lots of activity are active in communities with high populations, and may act as bridges between multiple communities with high activity and high population. However, the three metrics are not guaranteed to correlate and should be measured separately.

\begin{figure}[hbt]
    \centering
    \includegraphics[width=\linewidth]{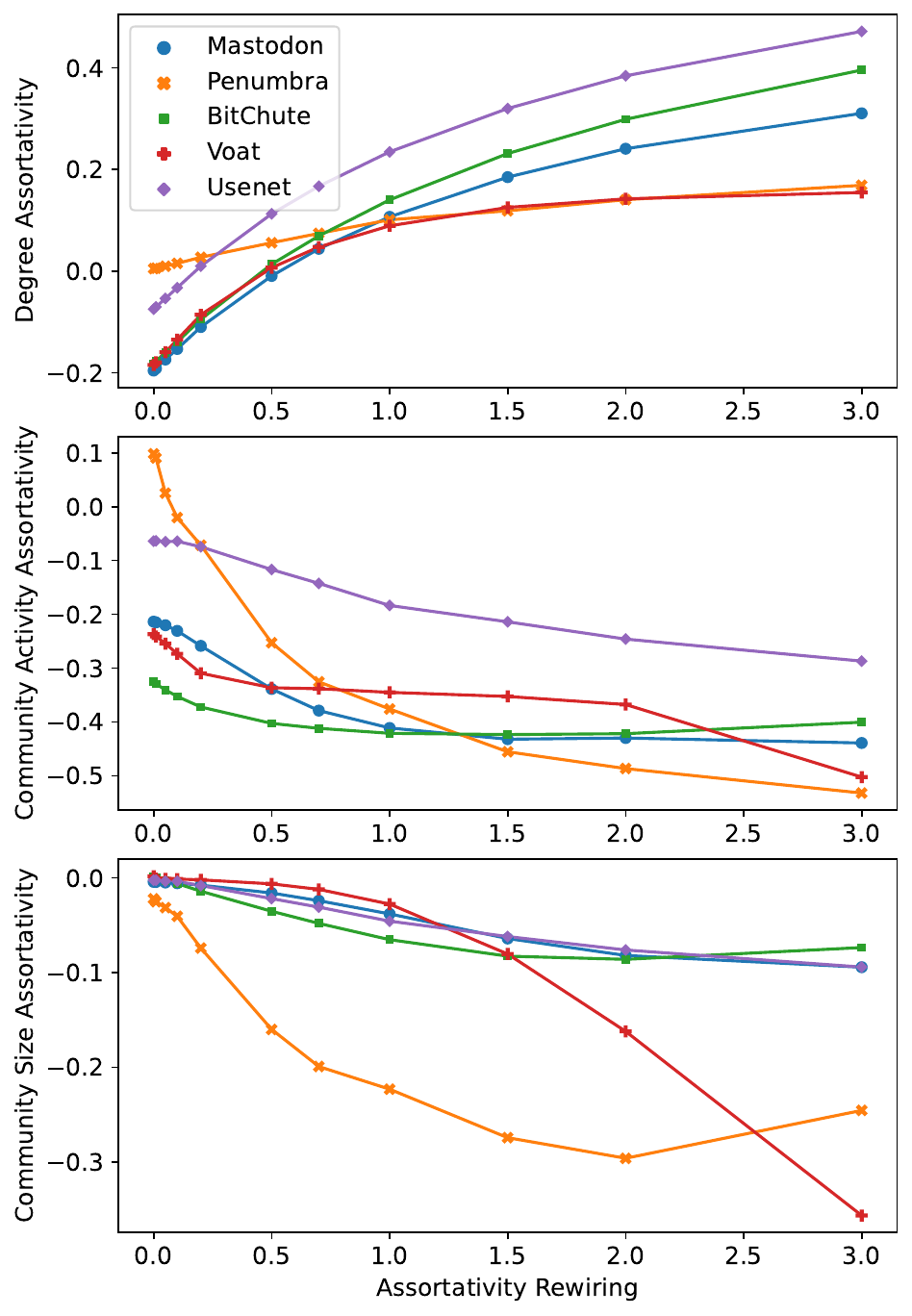}
    \caption{Rewiring to increase user-community degree assortativity (top) decreases the projected community-community degree assortativity (middle) and community-community population assortativity (bottom).}
    \label{fig:assortivity_user_vs_community}
\end{figure}

While rewiring to promote user-community degree assortativity, we also plotted the changes in community-community degree assortativity, shown in \cref{fig:assortivity_user_vs_community}. Strikingly, the community assortativity \textit{decreases} as we rewire to promote user assortativity. This is because as we rewire edges to focus user connections on the largest communities we implicitly decrease the number of edges between communities. This also matches the changes in disruption in \cref{fig:assortativity_auc}: increasing assortativity may reconnect large and insular communities with the rest of the network, briefly increasing their influence, but continued assortativity rewiring also cuts bridges to and between smaller communities, yielding a sparse network that is far less centralized.

\begin{figure}[hbt]
    \centering
    \includegraphics[width=\linewidth]{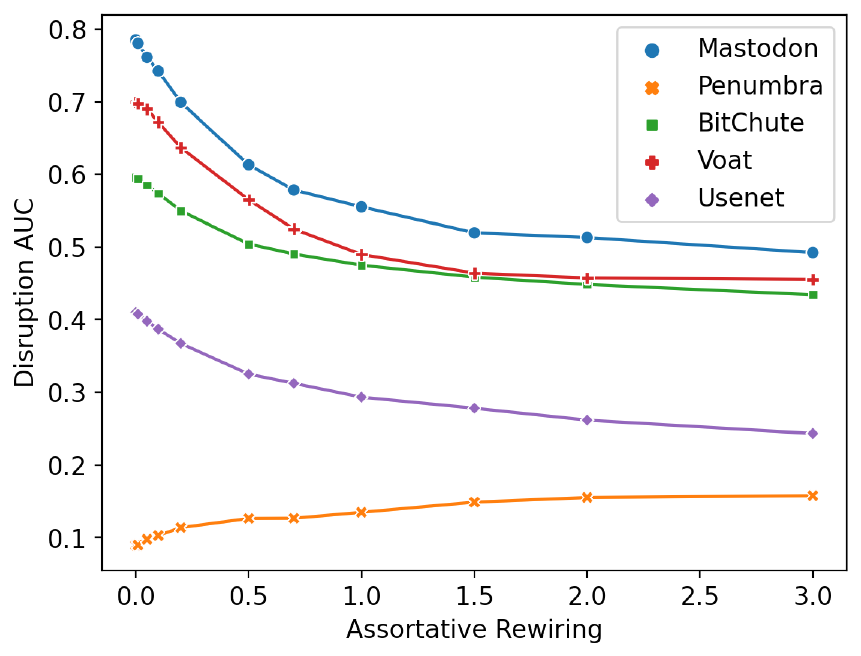}
    \caption{Rewiring networks to decrease user-community degree assortativity also typically decreases disruption when large communities are removed. However, for very sparse networks like the Penumbra, and perturbation, including rewiring to decrease assortativity, increases community inter-connection and so increases the influence of large communities.}
    \label{fig:disassortative_auc}
\end{figure}

\begin{figure}[hbt]
    \centering
    \includegraphics[width=\linewidth]{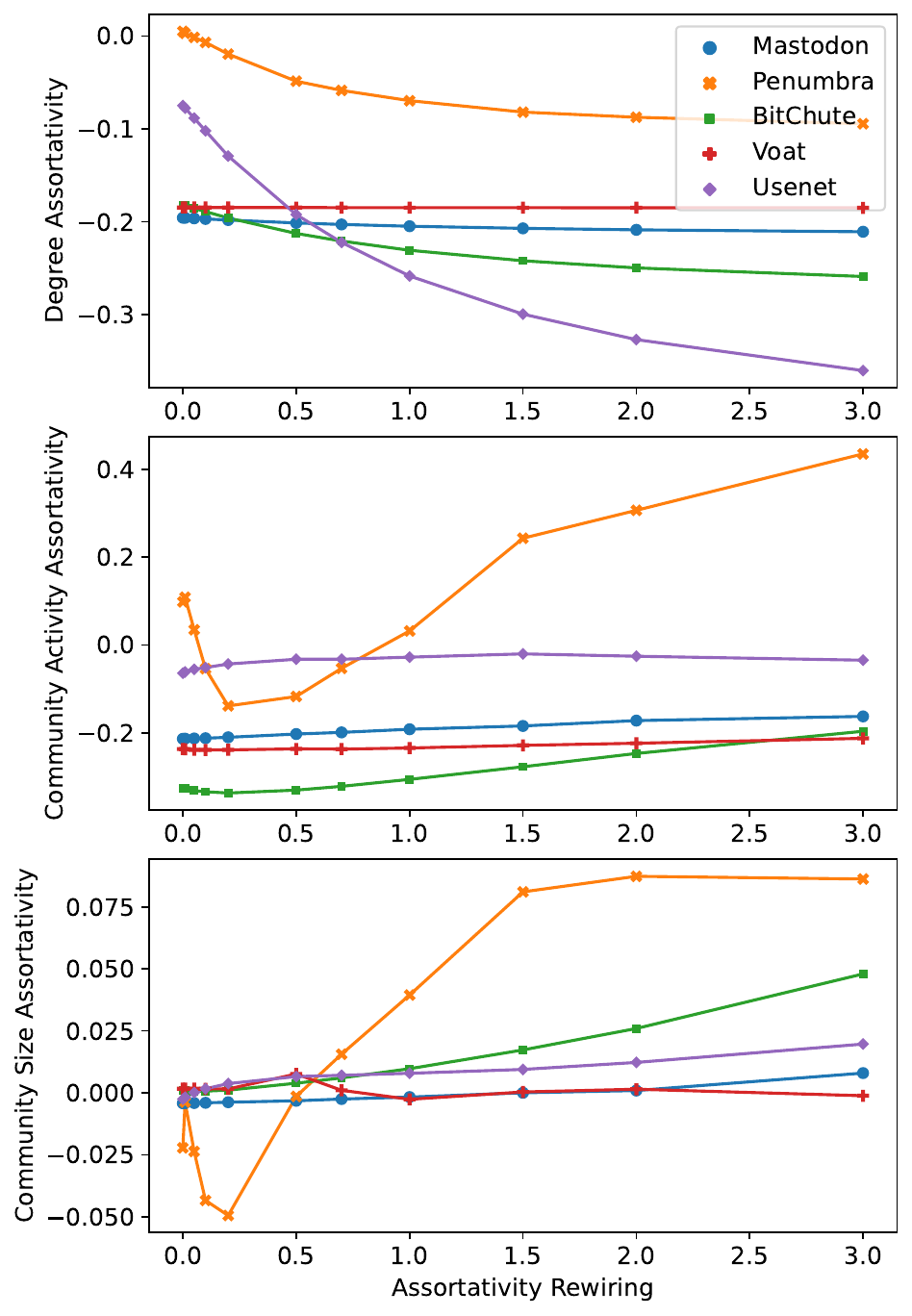}
    \caption{Rewiring to decrease user-community degree assortativity (top) has a small impact on the community-community projected degree assortativity (middle) and projected population assortativity (bottom), except for in the Penumbra; in this sparse network, rewiring first destroys the few active inter-community bridges, then radically increases the interconnectivity of communities.}
    \label{fig:disassortivity_user_vs_community}
\end{figure}

To further explore the relationship between these types of assortativity, we also rewired networks in the reverse direction: for randomly selected pairs of edges, we rewired those edges to \textit{decrease} user to community activity assortativity. We have plotted the change in disruption curves (\cref{fig:disassortative_auc}) and correlation between assortativity metrics (\cref{fig:disassortivity_user_vs_community}). In most networks, decreasing activity assortativity lowers centralization, although the effect diminishes as the network topology more closely approximates a random network. The one exception is the Penumbra; this network has such sparse inter-community connections that any perturbation of edges increases the cross-community links and therefore \textit{increases} centralization.

\section{Cumulative Impact on Giant Component Size} \label{sec:giant_components}

\begin{figure}[thb]
    \centering
    \includegraphics[width=\linewidth]{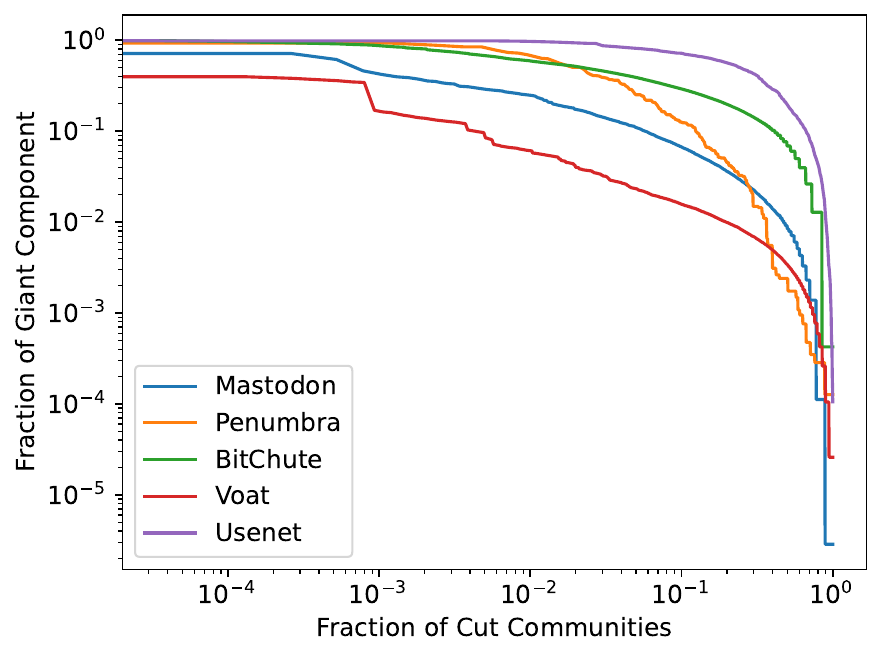}
    \caption{The giant component shrinks as communities are pruned from largest to smallest, indicating both the size of a community and whether it was part of the giant component before pruning. However, this boolean inclusion does not account for how well-integrated the community was among its peers. The y-axis is normalized as a fraction of the un-pruned giant component size, such that ``0.5" indicates the giant component is half the size of the original.}
    \label{fig:real_giant_component}
\end{figure}

Some readers may be interested in how removing large communities influences the giant component size on each network. This is closely related to the cumulative population size in the top sub-plots of \cref{fig:real_networks_size_comparison} and \cref{fig:toy_networks_size_comparison}. Intuition suggests that the size of the giant component will be inversely proportional to the number of cumulative communities removed; as more large communities are pruned, the giant component should shrink. This relationship holds so long as the remaining communities are interlinked, but falters once a ``bridge" community is removed and the giant component splinters. Therefore, sparsely connected networks where bridges are more prominent will have a chaotic giant component size, while more densely connected networks will present a smooth curve until most communities are pruned. This relationship is illustrated in \cref{fig:real_giant_component}. Most curves are smooth until the tail of the distribution, with two notable exceptions: Voat's giant component changes once the largest insular communities are removed (see \cref{fig:voat_render}), and the Penumbra's curve is much ``spikier" as a result of its highly sparse structure.

Measuring the change in giant component size captures some of the same features as our disruption metric. In particular, removing large insular communities may not change the giant component size if the community is completely isolated from the giant component, so this captures some aspect of both the size and topological role of a community. However, the impact of a community is boolean: if it touches the giant component, then removing the community will shrink the giant component by the size of that community. There is no distinction between a minimally integrated and tightly integrated community. Measuring the impact of a community in terms of fraction of edges severed, rather than component vertex size, offers finer insight into the interplay between size distribution and network structure.

\section{Comparison to Network Bottlenecking} \label{sec:cheeger}

The Cheeger number \cite{cheeger} is a single-valued metric representing how large of a ``bottleneck" inhibits conductance across a graph. It is typically written as:

\vspace{2em}
\begin{equation}
    h(G) = \min \left\{
        \frac{
                \eqnmarkbox[NavyBlue]{cheeger_crossedges}{|\partial A|}
            }{
                \eqnmarkbox[OliveGreen]{cheeger_alledges}{|A|}
            }
        : \eqnmarkbox[WildStrawberry]{cheeger_subset}{A \subseteq V(G)}, 
        \eqnmarkbox[Plum]{cheeger_bounds}{0 < |A| \leq \frac{1}{2} |V(G)|}
    \right\} 
\end{equation}
\annotate[yshift=1.2em]{above}{cheeger_crossedges}{Edges crossing the boundary of A}
\annotate[yshift=-0.2em]{below}{cheeger_alledges}{All edges in+across A}
\annotate[yshift=0.8em]{above}{cheeger_subset}{A is a subset of vertices of G}
\annotate[yshift=-2em]{below,left}{cheeger_bounds}{A contains at most half of all vertices}
\vspace{2em}

Our measurement of how much a community influences a larger population, and the Cheeger measurement of whether a community is a ``bottleneck" bear some conceptual similarities. Therefore, we compare our metric to the Cheeger number in two ways. First, we create a ``local Cheeger number," following an identical equation $\frac{|\partial A|}{|A|}$, but where $A$ is defined as the set of communities we are pruning, rather than via a global search. Second, we estimate bounds on the global Cheeger value of the graph. Since evaluating the graph conductance of all possible subsets of vertices is an NP-hard problem \cite{kaibel2004expansion}, it is impractical to directly measure the Cheeger constant on most large graphs. Fortunately, the Cheeger inequality offers upper and lower bounds on the Cheeger number based on the second eigenvalue of the normalized Laplacian of the adjacency matrix of G as follows:

$$\lambda_2/2 \leq h(G) \leq \sqrt{2\lambda_2}$$

\begin{figure}[hbt!]
    \centering
    \includegraphics[width=\linewidth]{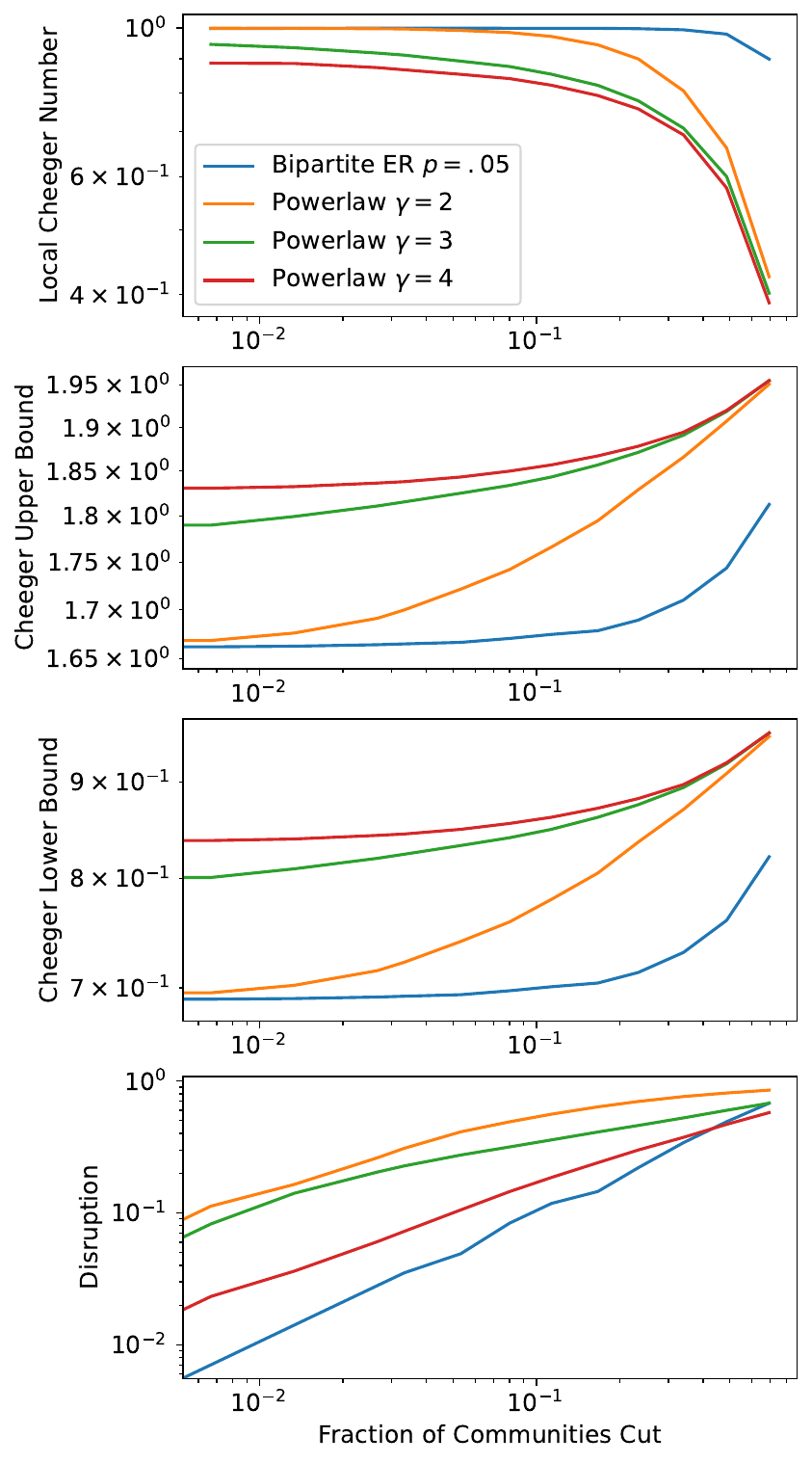}
    \caption{Our network disruption metric bears some conceptual similarity to network bottlenecks, but neither a ``local" Cheeger value measuring the bottleneck effect of removed communities (top) nor upper- and lower-bound estimates of the global Cheeger number describe the same trends.}
    \label{fig:cheeger}
    \vspace*{5.6cm} 
\end{figure}

Since they are sparse, these bounds can be calculated even on large real-world datasets. 
Unfortunately, in our tests the bounds are quite wide (see \cref{fig:cheeger}), limiting the utility of this approximation. We have plotted a comparison of the ``local" Cheeger number, bounds of the global Cheeger number, and our disruption metric, for a variety of simulated networks.

\printbibliography[heading=subbibliography]
\end{refsection}

\end{document}